\documentclass[twocolumn,showpacs,preprintnumbers,amsmath,amssymb]{revtex4}
\usepackage{graphicx}
\usepackage{dcolumn}
\usepackage{bm}

\begin{document}

\title{Effect of frequency detuning on pulse propagation in one-dimensional photonic crystal with a dense resonant medium: application to optical logic}

\author{Denis V. Novitsky}
 \email{dvnovitsky@tut.by}
\affiliation{%
B.I. Stepanov Institute of Physics, National Academy of Sciences of
Belarus, \\ Nezavisimosti~Avenue~68, 220072 Minsk, Belarus.
}%

\begin{abstract}
We consider propagation of light pulses detuned from the atomic
resonance in a dense two-level medium and photonic structures with
it. The large density of the medium is important to decrease spatial
scale of such nonlinear effects as pulse compression, though it does
not provide any fundamentally new phenomena as compared to dilute
media. Frequency detuning decreases the effectivity of such
nonlinear phenomena as pulse compression and dispersion spreading
compensation as well. We propose simple logic gates based on
interaction between two pulses in one-dimensional nonlinear photonic
crystal. It turned out, that frequency detuning is necessary to
obtain ultrafast AND gate, while OR and NOT gates can be realized in
the system without detuning.
\end{abstract}

\maketitle

\section{\label{intro}Introduction}

A dense resonant medium is a model of two-level medium which takes
into account near dipole-dipole (NDD) interactions between atoms.
NDD interactions described by local field correction lead to
appearance of intrinsic optical bistability \cite{Hopf, Malyshev,
Afan98} or change the characteristics of such effects as optical
switching \cite{Cren92, Scalora}, soliton formation \cite{Bowd91,
Afan02}, etc.

NDD interactions effects are often treated as a result of nonlinear
dynamic shift of atomic resonance from the frequency of input
radiation. The efficiency of this process is governed by the value
$b=4 \pi \mu^2 C/3 \hbar \gamma_2$, where $\mu$ is the transition
dipole moment, $\hbar$ is the Planck constant, $\gamma_2=1/T_2$ is
the rate of transverse relaxation. The quantity $b$ is usually
referred to as a constant of NDD interactions and is limited by the
volume density of two-level atoms $C$. Though the magnitude of $b$
is extremely important, the static detuning $\Delta\omega$ of light
frequency from atomic resonance should be taken into account as
well, as it determines existence regions for a number of phenomena
in dense resonant media. For example, local intrinsic optical
bistability (IOB) occurs in thin films if the parameter of NDD
interactions satisfies the condition \cite{NovJOSA}
\begin{eqnarray}
b > 2 \delta' + 6 (1+\delta'^2)^{1/3} {\rm Re} (\delta'+i)^{1/3},
\label{iobbound}
\end{eqnarray}
so that its minimal value required for IOB amounts to $4$
\cite{Friedberg} at the detuning $\delta'=\Delta\omega/\gamma_2=-1$.
Another effect is connected with appearance of the so-called
incoherent solitons which require static detuning to be negative
\cite{Afan02}. On the other hand, detuning should not be too large:
if it is much greater than NDD interaction parameter, the resonant
medium behaves as a dilute one \cite{Cren96}. This is the reason why
it is often assumed to be zero in the problem of ultrashort pulse
propagation through the dense resonant medium.

Since in existing literature there is no consistent analysis of
static frequency detuning influence on pulse propagation in dense
resonant medium, we try to fill up this gap in the present paper. We
consider coherent pulse transformation in a dense resonant medium
with realistic values of NDD interaction parameter and apply nonzero
detunings which are large enough to cause significant effects.
Another aspect of the present research, which is of particular
interest, is connected with pulse propagation in nonuniform dense
medium. Periodic variations of its linear parameters form the
one-dimensional photonic crystal. The problem of enhancement of
nonlinear pulse transformation effectiveness and pulse controlling
by another pulse in such systems is also investigated for the case
of nonzero detuning.

The paper contains three sections. In Sec. \ref{DRM} the main
equations and parameters are considered, as well as the problem of
medium density. The peculiarities of pulse compression in a dense
resonant medium and photonic crystal with it in the case, when
frequency detuning is present, are discussed in Sec. \ref{compr}.
Sec. \ref{gates} is devoted to a perspective from practical
viewpoint result connected with interaction between two pulses --
realization of optical logic elements on the basis of nonlinear
photonic crystal. As it is demonstrated, detuning of nonlinear
medium from resonance plays crucial role allowing to make the
operation of logic gates faster and more stable.

\section{\label{DRM}Dense resonant medium: Importance of density}

Dynamics of pulse propagation in dense resonant medium are described
in semiclassical approach by the system of generalized Maxwell-Bloch
equations \cite{Bowd93, Cren96}:
\begin{eqnarray}
\frac{dP}{dt}&=&\frac{i \mu}{\hbar} E N + i P \left(\Delta \omega +
\frac{4 \pi \mu^2 C}{3 \hbar} N \right) - \gamma_2 P,
\label{dPdt} \\
\frac{dN}{dt}&=&2 \frac{i \mu}{\hbar} \left(E^* P - P^* E \right)
-\gamma_1 (N - 1), \label{dNdt} \\
\frac{\partial^2 \Sigma}{\partial z^2}&-&\frac{1}{c^2}
\frac{\partial^2 \varepsilon_{bg} \Sigma}{\partial t^2} = \frac{4
\pi}{c^2} \frac{\partial^2 P_{nl}}{\partial t^2}, \label{Max}
\end{eqnarray}
where $N$ is the population difference, $P$ is the microscopic
(atomic) polarization; $\gamma_1=1/T_1$ is the rate of longitudinal
relaxation; $c$ is the light speed in vacuum; $k=\omega/c$ is the
wavenumber, and $\varepsilon_{bg}$ is the background dielectric
permittivity (linear and dispersionless). Macroscopic electric field
$\Sigma$ is expressed via its amplitude $E$ as $\Sigma=E
\exp[-i(\omega t-kz)]$; similarly for macroscopic nonlinear
polarization we have $P_{nl}=\mu CP \exp[-i(\omega t-kz)]$.

The system of equations (\ref{dPdt}-\ref{Max}) can be written in
dimensionless form as follows:
\begin{eqnarray}
\frac{dP}{d\tau}&=&i \tilde\Omega N + i P (\tilde\delta+\epsilon N) - \tilde\gamma_2 P, \label{dPdtau} \\
\frac{dN}{d\tau}&=&2i (\tilde\Omega^* P - P^* \tilde\Omega) -
\tilde\gamma_1 (N-1), \label{dNdtau} \\
\frac{\partial^2 \tilde\Omega}{\partial \xi^2}&-&\varepsilon_{bg}
\frac{\partial^2 \tilde\Omega}{\partial \tau^2}+2i \frac{\partial
\tilde\Omega}{\partial \xi}+2i \varepsilon_{bg} \frac{\partial
\tilde\Omega}{\partial \tau}+(\varepsilon_{bg}-1)
\tilde\Omega \nonumber \\
&&=3 \epsilon \left(\frac{\partial^2 P}{\partial \tau^2}-2i
\frac{\partial P}{\partial \tau}-P\right), \label{Maxdl}
\end{eqnarray}
$\tau=\omega t$ and $\xi=kz$ are dimensionless time and space
arguments; $\tilde\Omega=(\mu/\hbar\omega)E$ is the dimensionless
amplitude of electric field; $\tilde\delta=\Delta\omega/\omega$ is
the normalized static frequency detuning; $\epsilon=b
\tilde\gamma_2$; $\tilde\gamma_j=\gamma_j/\omega$, $j=1, 2$; and
$\varepsilon_{bg}$ is the background dielectric permittivity.

In the present paper the input signals are assumed to be coherent
Gaussian pulses with amplitude
$\tilde\Omega=\tilde\Omega_0\exp(-t^2/2t_p^2)$, i.e., for pulse
duration $t_p$, the inequality is valid: $t_p<<T_2<T_1$. In our
calculations the parameters of the medium and pulse are as follows:
$T_1=1000$ ps and $T_2=100$ ps, pulse duration $t_p=30$ fs and the
wavelength $\lambda=0.5$ $\mu$m. Pulse amplitude can be measured in
the units of $\Delta\Omega_T=\lambda/2 \sqrt{2\pi}c t_p$, which is a
natural parameter connected with optical switching \cite{NovPRA}. A
pulse with $\tilde\Omega_0=\Delta\Omega_T$ inverts the medium and
then returns it exactly to the ground state, so that it can be
treated as a $2\pi$-pulse. Periodic variation of background
dielectric constant $\varepsilon_{bg}$ describes the situation of
one-dimensional nonlinear photonic crystal. In order to accurately
take into account processes connected with dispersion, the numerical
simulation of the system of equations (\ref{dPdtau}-\ref{Maxdl}) is
implemented on the basis of approach developed in \cite{Cren96,
NovPRA}.

The additional nonlinear term with $\epsilon N$ in Eq.
(\ref{dPdtau}), which couples polarization and population
difference, is responsible for near dipole-dipole interactions. As
it was shown in Ref. \cite{NovPRA}, influence of this term on
coherent pulse propagation (in contrast to consideration in
stationary regime) is negligible at realistic values of the
parameter $b$ which is usually limited to several units (at best,
$b=10$). Though it will be the case further, the dense resonant
medium has a very important advantage over dilute media. The density
of the medium is present in the right hand side of the wave equation
(\ref{Maxdl}) due to the macroscopic polarization definition (it
measures the dipole moment of macroscopic volume) and, hence,
influences on pulse propagation. This influence is not physically
connected with NDD interactions (formally, there is just the same
combination of parameters $\epsilon$) and determines spatial scale
of nonlinear effects in the resonant two-level medium. High values
of the density and, consequently, of the constant $b$ allow to
obtain such phenomena as pulse compression on significantly reduced
distances. Indeed, the distance of optimal compression (limited by
the processes of diffraction, dispersion and energy absorbtion)
\cite{NovPRA} approximately equals $1000\lambda$ for $b=1$, while
for $b=10$ it is only about $100\lambda$ (for the pulse with
$\tilde\Omega=1.5 \Delta\Omega_T$). Thus, the large density of
resonant medium is important to increase effectiveness of nonlinear
effects and compactness of their applications due to right hand side
of Eq. (\ref{Maxdl}), though it does not provide any fundamentally
new phenomena connected with NDD interactions term in Eq.
(\ref{dPdtau}).

The previous results \cite{NovPRA} were obtained for the case of
exact resonance, i.e. detuning $\tilde\delta$ of the field frequency
from atomic resonance was zero. Further we consider influence of
nonzero detuning on pulse compression in the dense resonant medium
and photonic crystal containing it.

\section{\label{compr}Effect of frequency detuning on pulse compression}

\begin{figure}[t]
\centering \includegraphics[scale=0.85, clip=]{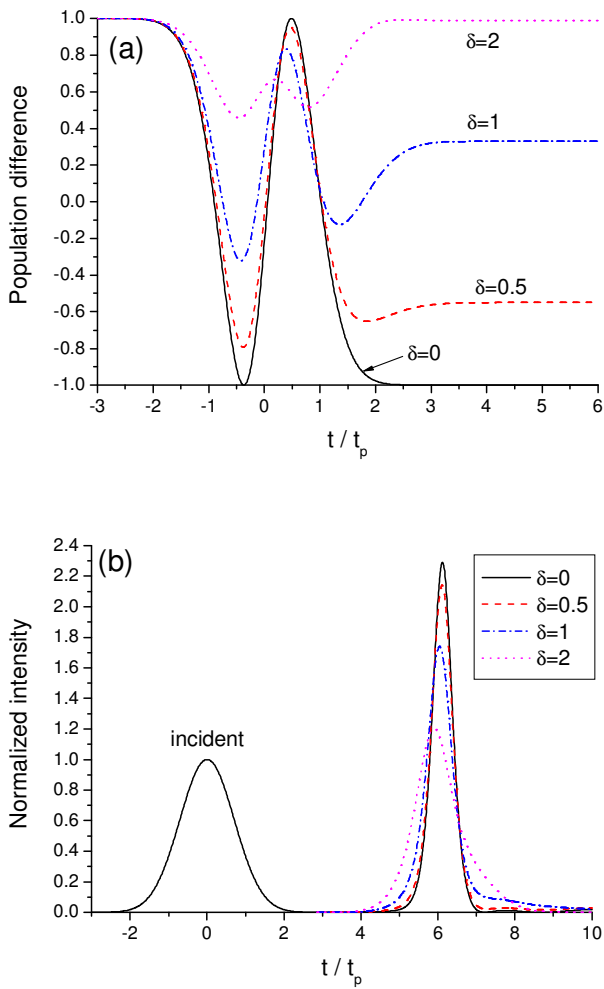}
\caption{(Color online) (a) Dynamics of population difference on the
entrance of the layer of the dense resonant medium after pulse
passage and (b) transmitted pulse form transformation for different
values of frequency detuning $\delta$. The amplitude of pulse is
$\Omega_0=1.5$. Layer thickness is $L=100\lambda$.} \label{fig1}
\end{figure}

\begin{figure}[t]
\centering \includegraphics[scale=0.85, clip=]{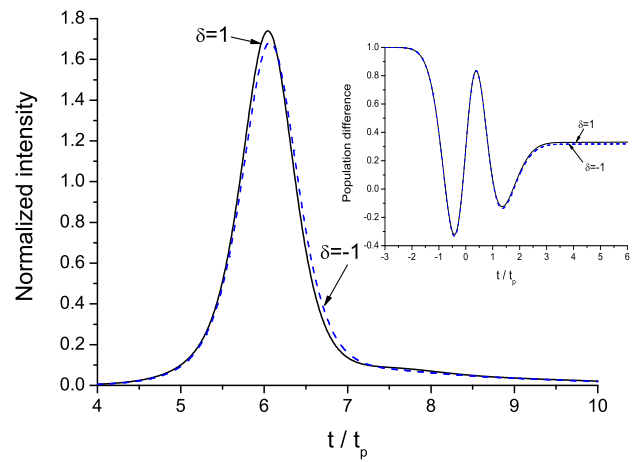}
\caption{(Color online) Comparison of transmitted pulse forms for
differnt signs of frequency detuning $\delta$. The amplitude of
pulse is $\Omega_0=1.5$. Layer thickness is $L=100\lambda$. The
inset shows corresponding dynamics of population difference.}
\label{fig2}
\end{figure}

Let us consider propagation of short coherent ($t_p<<T_2$) pulse in
a dense resonant medium with NDD interaction constant $b=10$ and
background permittivity $\varepsilon_{bg}=1$ (other parameters are
stated in the previous section), and pulse intensity is high enough
to obtain inversion of medium, i.e.
$\tilde\Omega_0\sim\Delta\Omega_T$. It is easy to assume that
influence of frequency detuning $\tilde\delta$ becomes significant
when it is comparable with this dimensionless value
$\Delta\Omega_T$, too. This is proved by Fig. \ref{fig1} (here and
further pulse amplitude and detuning of the medium are measured in
the units of $\Delta\Omega_T$, i.e.
$\delta=\tilde\delta/\Delta\Omega_T$ and
$\Omega=\tilde\Omega/\Delta\Omega_T$). It is seen that, when
increasing detuning from the resonance, the final quasistationary
level of population difference is shifted from the fully inverted
state (if detuning is absent) almost to the ground state (for
$\delta=2$). At the same time the effectivity of pulse compression
is decreasing (see Fig. \ref{fig1}b). The sign of frequency detuning
almost does not affect the behavior of pulse in the dense resonant
medium, according to the comparison of results for $\delta=1$ and
$\delta=-1$ demonstrated in Fig. \ref{fig2}. Note that the detuning
$|\delta|=1$ corresponds (for parameters used) to the frequency
shift of about $\Delta\omega\approx0.011\omega$.

\begin{figure}[t]
\centering \includegraphics[scale=0.85, clip=]{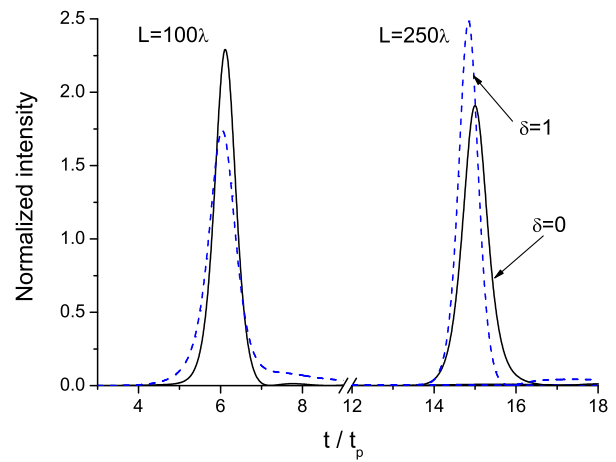}
\caption{(Color online) Comparison of transmitted pulse forms for
different frequency detunings $\delta$ at different layer
thicknesses. The amplitude of pulse is $\Omega_0=1.5$.} \label{fig3}
\end{figure}

Nevertheless, the same or even higher pulse compression can be
reached for more thick layers. This means that, for $\delta\sim1$,
longer distances are needed to provide effective nonlinear
interaction of pulse with the dense resonant medium and, hence, the
distance of optimal compression (limited by diffraction and
dispersion) is increasing, as well. For example, it amounts to about
$100\lambda$ for $\delta=0$, while for $\delta=1$ it is
approximately $250\lambda$ (see Fig. \ref{fig3}).

\begin{figure}[t]
\centering \includegraphics[scale=0.85, clip=]{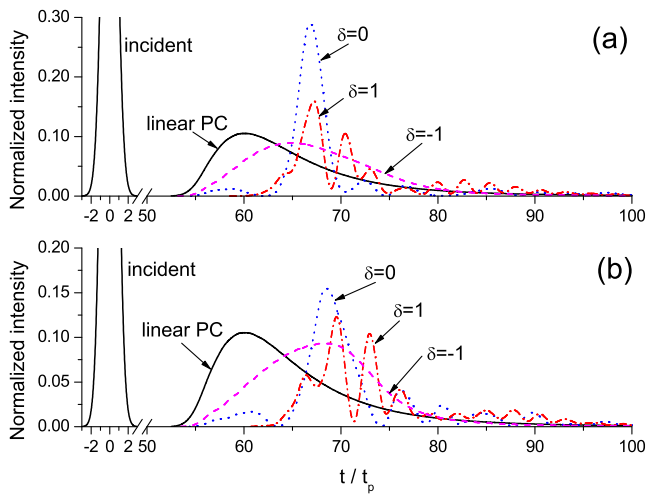}
\caption{(Color online) The forms of pulses transmitted through the
photonic crystal with (a) nonlinear layers $d_1$, (b) both nonlinear
layers $d_1$ and $d_2$ at different values of frequency detuning
$\delta$. The number of periods is 250, the thicknesses $d_1=0.4$,
$d_2=0.13$ $\mu$m, background refractive indices $n_1=1$, $n_2=3.5$.
The amplitude of pulse is $\Omega_0=1.5$.} \label{fig4}
\end{figure}

Another negative influence of frequency detunings is connected with
the effect of compensation of dispersion spreading in nonlinear
photonic crystals found in Ref. \cite{NovPRA}. While pulse spreads
rapidly as it propagates in linear photonic crystal (which is an
element with high level of dispersion), nonlinear compression allows
to compensate this widening. Fig. \ref{fig4} shows that this
property is significantly reduced for $\delta=1$ (several
low-intensive pulses appear instead of a single high-intensive one)
and, especially, for $\delta=-1$ (one broadened low-intensive
pulse).

Thus, off-resonant interaction of a single light pulse with high
enough detunings $\delta\sim1$ with the dense two-level medium
results in decrease of efficiency of such important processes as
pulse compression and dispersion spreading compensation. However,
frequency detuning is turned out to be useful or even necessary
condition for two-pulse interactions and, in particular, for
realization of optical logic elements on the basis of nonlinear
photonic structure.

\section{\label{gates}Optical logic gates}

In Ref. \cite{NovPRA} it was shown that pulse intensity can be
controlled with another pulse in the scheme of two pulses
copropagating in a resonant two-level medium. Moreover, using of
photonic crystal allows to significantly enhance the efficiency of
this process. This phenomenon is a basic idea for optical logic
elements to be considered.

Possibility of optical logic operations is one of the most
prospective results in nonlinear optics of photonic crystals.
Usually logic gates are realized by using two- and three-dimensional
photonic band-gap structures with defects which form waveguides and
nonlinear resonators (see, e.g., \cite{Asakawa, Andalib} and
references there). Such systems must have, at least, two inputs for
initial signals and one output for resulting one. It seems to be
much more simple to use one-dimensional structures with the same
entry for both input signals. An example of one-dimensional logic
gates on the basis of nonlinear semiconductor heterostructure was
given in \cite{Nefedov}. The possibility of one-dimensional logic
elements based on photonic crystal with dense resonant medium is to
be discussed further.

The important property of the gates proposed is that the same
nonlinear structure can serve for different logic gates, depending
on the intensity of input signals. (Similar property was used in
\cite{Nefedov} to demonstrate optical elements in stationary
regime.) The first pulse changes the characteristics of nonlinear
medium, as it propagates in it, while the second one interacts with
this modified medium. This effect can be made more sharp by using
photonic crystal. In this case specific reflection and transmission
properties of this structure become essential.

\begin{figure}[t!]
\centering \includegraphics[scale=0.85, clip=]{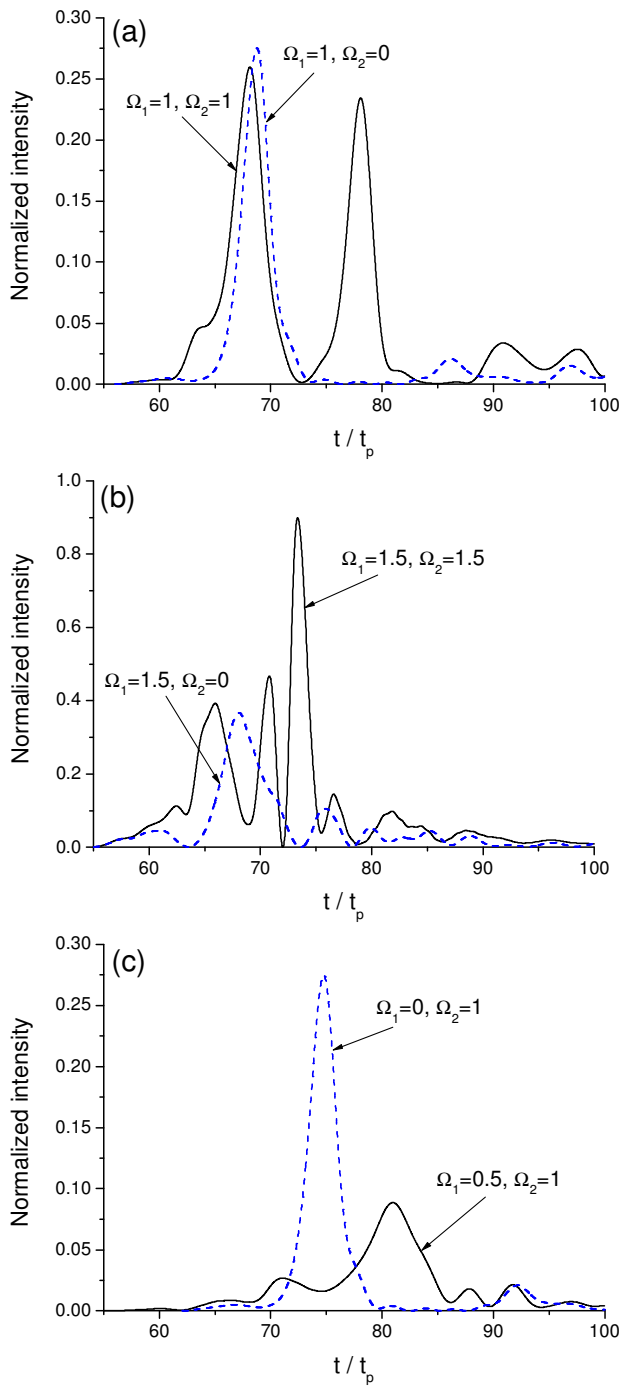}
\caption{(Color online) Shapes of transmitted pulses for different
input signals corresponding to (a) OR gate, (b) AND gate, (c) NOT
gate. The number of periods is $250$, the thicknesses of the layers
$d_1=0.4$ $\mu$m, $d_2=0.1301$ $\mu$m, background refractive indices
$n_1=1$, $n_2=3.5$. Frequency detuning is $\delta=0$.} \label{fig5}
\end{figure}

First, let us consider the case $\delta=0$. Fig. \ref{fig5} shows
the example of numerical simulation of Eqs.
(\ref{dPdtau}-\ref{Maxdl}) for the photonic crystal with
reflectivity $R=0.469$ at the main wavelength $\lambda$. Different
logic operations can be achieved for different levels of input
intensities. OR gate (disjunction) corresponds to the case when, in
both cases of one and two pulses on the entrance, one obtains
approximately the same peak intensity on the output. This situation
occurs for the input pulses with the intensities
$\Omega_1=\Omega_2=1$ (in units of $\Delta\Omega_T$) and is
demonstrated in Fig. \ref{fig5}a. The results for transmission of
these pulses are summed up in the section $1$ of Table
\ref{tab:table1}. It is possible to say that this section represents
a truth table for the OR element considered.

\begin{table}
\caption{\label{tab:table1}Table of output peak intensities (truth
tables)}
\begin{center}
\begin{tabular}{l|ccc|cc}
\hline
Gate at $\delta=0$ & $\Omega_1$ & $\Omega_2$ & $\delta=0$ & $\delta=1$ & $\delta=-1$ \\
\hline
1. OR& 0 & 0 & 0 & 0 & 0\\
&1 & 0 & 0.28 & 0.14 & 0.07\\
&0 & 1 & 0.28 & 0.14 & 0.07\\
&1 & 1 & 0.26 & 0.19 & 0.37\\
\hline
2. AND &1.5 & 0 & 0.37 & 0.27 & 0.23\\
&0 & 1.5 & 0.37 & 0.27 & 0.23\\
&1.5 & 1.5 & 0.90 & 0.84 & 1.01\\
\hline
3. NOT & 0 & 1 & 0.28 & 0.14 & 0.07\\
&0.5 & 1 & 0.09 & 0.11 & 0.19\\
\hline
4. OR &1.5 & 0 & 0.37 & 0.27 & 0.23\\
&0 & 1 & 0.28 & 0.14 & 0.07\\
&1.5 & 1 & 0.33 & 0.60 & 0.70\\
\hline
5. AND &1 & 0 & 0.28 & 0.14 & 0.07\\
&0 & 1.5 & 0.37 & 0.27 & 0.23\\
&1 & 1.5 & 0.85 & 0.70 & 0.70\\
\hline
\end{tabular}
\end{center}
\end{table}

AND gate (conjunction) occurs when, as a result of nonlinear
interaction of two pulses, the intensity of the transmitted one is
much greater than in the case of a single pulse. This situation
takes place for $\Omega_1=\Omega_2=1.5$ and is shown in Fig.
\ref{fig5}b. The corresponding truth table (Table \ref{tab:table1},
section $2$) shows that the state when both signals are present
(upper or AND-state) gives almost $2.5$-times higher peak output
intensity than the state when one of them is absent (lower state).

Finally, NOT operation can be realized as follows. Propagation of a
single pulse $\Omega_2=1$ gives the output peak intensity of about
$0.28$ (see Fig. \ref{fig5}c), so that it can be considered as
ON-state. By using another pulse of $\Omega_1=0.5$, the output
intensity is dramatically decreased to $0.09$, i.e. more than by a
factor of three. This low level of transmitted intensity corresponds
to the OFF-state. This effect of switching between ON- and
OFF-states forms the basis of NOT gate. Its truth table is shown in
section $3$ of Table \ref{tab:table1}.

As it is seen from the Table \ref{tab:table1}, the gates also can be
realized for other combinations of input pulse intensities. For
example, OR gate occurs when $\Omega_1=1.5$, $\Omega_2=1$, while for
the reverse case ($\Omega_1=1$, $\Omega_2=1.5$) one obtains AND
operation.

As it was mentioned, result of interaction between pulses strongly
depends on the properties of photonic band-gap structure and, in
particular, on its reflectivity. For example, if we take photonic
crystal with large linear reflectivity, namely $R=0.825$ (it can be
made by changing the thickness $d_2$ to $0.1274$ $\mu$m), logic
operations described above become impossible. For the pulses with
the amplitude $\Omega_1=\Omega_2=1.5$ (it was the AND gate in Fig.
\ref{fig5}b) the results of pulse propagation are as follows: output
intensity of $0.42$ for a single pulse and $0.53$ for two pulses,
i.e. there is no pronounced AND-state. The contrast between states
in other cases gets less sharp, too. Thus, realization of the
optical logic operations is the result of consistent change of pulse
intensities and nonlinear photonic crystal properties: during
propagation, pulse changes the characteristics of nonlinear photonic
band-gap structure, which, in turn, has an influence on the pulse
intensities. This mutual change leads to complexity of the effect of
logic operations and nonobviousness of its appearance at certain
parameters of the system.

Now let us consider the results of simulations for the case
$|\delta|=1$ which are represented in Table \ref{tab:table1} (two
last columns). It is seen that frequency detuning increases the
contrast between lower and upper states of the AND gate. Moreover,
OR and NOT gates (at $\delta=0$) become less pronounced or even
transform to the AND gate when the detuning is present. This is
especially typical for negative detuning $\delta=-1$. Thus, the
system with detuning can be used to obtain high-contrast AND
operation, while OR and NOT gates are realized at $\delta=0$. It is
important, that AND gate with detuning has another advantage
connected with its bit rate.

\begin{figure}[t]
\centering \includegraphics[scale=0.85, clip=]{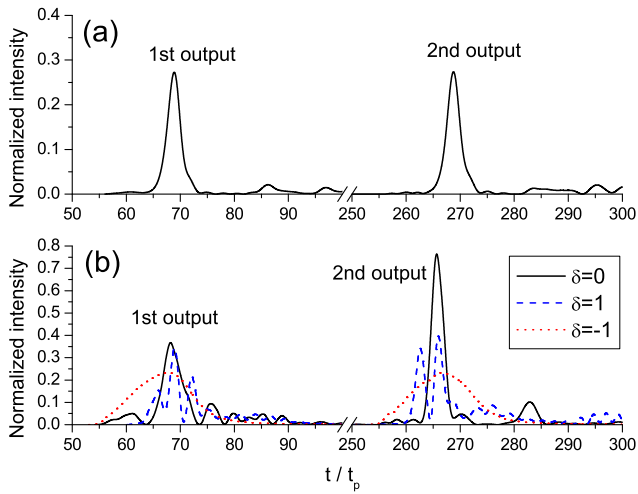}
\caption{(Color online) Shapes of transmitted pulses for two
sequential single-pulse inputs with time interval of $200 t_p$
between them. Input pulse intensities are (a) $\Omega_0=1$ at
$\delta=0$ (OR gate), (b) $\Omega_0=1.5$ at different $\delta$ (AND
gate). Other parameters are the same as in Fig. \ref{fig5}.}
\label{fig6}
\end{figure}

Since the logic elements described above are based on interaction
between one-by-one propagating pulses, a certain time interval
should separate sequential realizations of logic operations. This
period provides independent interaction of pulses with nonlinear
medium and characterizes a bit rate of the logic gates proposed. In
general, it is determined by the relaxation times of the resonant
medium. Then the bit rate can be calculated as $1/T_2$, i.e. about
$10$ Gbit/s. It is not high level, as compared with Kerr-type
structures (see, for example, \cite{Andalib}). Nevertheless, in some
cases this rate can be increased. Fig. \ref{fig6}a shows propagation
of two pulses with $\Omega_0=1$ used in OR gate at $\delta=0$. It is
seen that time interval of about $\Delta t=200 t_p$ between input
signals is enough for independent propagation of pulses and
operation of OR gate. For the parameters used in simulations we have
$t_p=30$ fs and $\Delta t=6$ ps, which is much less than relaxation
times of the resonant medium. Hence, the rate of logic gates
operation, which can be estimated as $1/\Delta t$, amounts to a
value of more than $160$ Gbit/s. Obviously, even greater bit rates
can be achieved for OR gate due to weak interactions between pulses
in this case.

However, this is not the case for AND gate at $\delta=0$.
Propagation of two single pulses with $\Omega_0=1.5$ separated by
the time interval of $\Delta t=200 t_p$ does not give two
independent low states of AND gate (Fig. \ref{fig6}b, solid line).
This means that pulses interact due to slow relaxation of the
medium. As it is seen from Fig. \ref{fig1}a, presence of frequency
detuning results in faster relaxation of the dense resonant medium.
This can be used to increase bit rate of AND gate. Indeed, the
interaction between pulses is significantly weaker for $\delta=1$
(dashed line in Fig. \ref{fig6}b) or even almost absent for
$\delta=-1$ (dotted line). Thus, frequency detunings $|\delta|\sim1$
are useful to realize ultrafast (more than $160$ Gbit/s) logic
operation (in particular, AND).

\section{\label{conc}Conclusion}

Summing up, propagation effects for light pulses with frequency
detuned from the atomic resonance in the dense two-level medium
significantly differs from those for resonant radiation. Such
important phenomenon as pulse compression shifts towards larger
distances, while compensation of dispersion spreading in photonic
crystal becomes less pronounced.

Special attention is attracted to possibitity of logic gates
realization in nonlinear photonic crystals. The type of logic
element is governed by the intensity of input pulses. It turned out
that stable and fast OR and NOT gates can be obtained in the system
with zero frequency detuning. Detuning becomes necessary in the case
of AND gate resulting in fast relaxation and increasing bit rate to
the level of more than $160$ Gbit/s.

Note that the recompense for simplicity and ultrafast operation of
the logic gates considered is high level of peak intensity of input
pulses required to observe the effects described above. While, for
stationary cases of logic elements \cite{Nefedov} or optical
switching \cite{NovJOSA}, input intensities are needed to be of
about units and tens kW/cm$^2$, in our case peak intensity reaches
values of the order of GW/cm$^2$. This situation seems to be typical
for usage of ultrashort light pulses. The reason is that
approximately the same energy as in stationary regime should be
transferred in much shorter time in pulse regime.

Can the effects described be really observed? In this paper we
investigated this possibility in principle, but not a particular
device. Nevertheless, the used characteristics of the dense resonant
medium (the relaxation times and the constant $b$), which is
responsible for nonlinear interaction, are typical for usually
discussed candidates -- dense gases and excitonic media
\cite{Afan98, Afan02, NovPRA}. At the same time, the parameters of
photonic crystal (background refractive indices, layer thicknesses,
number of periods) are not so important, because any changes in
reflectivity spectrum of the structure can be compensated, for
example, by adjusting the thickness of the layers. Thus, though the
structure considered is quite hypothetical, there is no reason to
say that it is not realistic.

\end{document}